\documentclass[12pt,preprint]{aastex}
\begin{document}
\title{On the Environmental Dependence of Galaxy Properties
Established by the Initial Cosmological Conditions}
\author{\sc Jounghun Lee}
\affil{Department of Physics and Astronomy, Seoul National University, 
Seoul 151-742, Korea}
\email{jounghun@astro.snu.ac.kr}

\begin{abstract}
We study theoretically how the initial cosmological conditions establish 
the dependence of galaxy properties on the environment. First, we adopt the 
linear tidal torque theory according to which the angular momentum of a 
proto-galaxy is generated at first order by the misalignment between the 
proto-galaxy inertia tensor and the local tidal tensor.  Then, we quantify 
analytically the degree of the misalignment between the two tensors, and show 
quantitatively that it increases as the density of the environment decreases. 
It implies that the proto-galaxies forming in the lower density regions 
should end up with having higher angular momentum than those in the higher 
density regions, which is consistent with recent numerical finding that the 
void and field galaxies have higher spin parameters than the cluster galaxies. 
Since the galaxy angular momentum plays a role of developing a disk-like 
structure and hindering the star-formation, our theoretical insight provides 
an answer to such  fundamental observational question as why the large void 
galaxies have young stellar populations and high specific star formation 
rate,  which was not explained by the previous morphology-density 
relation. 
\end{abstract} 

\section{INTRODUCTION}

Recent observations from the Sloan Digital Sky Survey \citep[SDSS,][]
{str-etal02} and the 2dF Galaxy Redshift Survey \citep[2dFGR,][]{col-etal01}
have made it possible to study the properties of galaxies from largest 
spectroscopic samples. Although it has been noted for long that 
the morphology and the specific star formation rate of galaxies depend on 
the density of environments \citep{dre80,pos-gel84,whi-etal93}, it is only in 
recent years that the environmental dependence of galaxy properties have been 
observationally confirmed as common and studied systematically through 
statistical analyses of these large samples \citep{lew-etal02,gom-etal03,
got-etal03,tan-etal04,roj-etal05}. 

Now that the environmental dependence of galaxy properties holds a crucial 
key to the galaxy formation, it is quite important and necessary to understand 
what physical mechanism establishes it.  A fundamental question that naturally 
arise in the study of the environmental dependence of galaxy properties is 
whether it is destined from the initial conditions or acquired during the 
evolutionary process. 

Previous studies were largely focused on the properties of those galaxies 
located in the dense environments like cluster cores, highlighting mainly 
the effect of the subsequent evolutionary process like hierarchical 
merging, galaxy-galaxy interactions, and so on \citep{dre80,pos-gel84,
whi-etal93,lew-etal02,tan-etal04}.  Yet, significant as its effect may be, 
the evolutionary process by itself is incapable of explaining the whole 
picture.  There are many aspects of the environmental variations of galaxy 
properties that cannot be explained by the effect of the evolutionary process. 
For example,  the analysis of SDSS data  has revealed recently that the 
large void galaxies have very young stellar populations and quite high 
specific  star formation rate \citep{roj-etal05}.  Since the rates of galaxy 
merging and interaction in voids are quite low, it is unlikely due to the 
effect of the evolutionary process.  Although it was claimed by
\citet{vog-etal04} that it is consistent with the predictions of the 
semi-analytic model of the galaxy formation \citep{ben-etal03},  
it has not been understood yet what caused the galaxies in voids to 
have such properties.  

What is generally suspected is that the initial cosmological conditions should 
be at least partly responsible for establishing the  environment-variation of 
galaxy properties \citep[e.g.,][]{tan-etal04}.   However, it is hard to 
investigate theoretically the effect of the initial cosmological conditions 
as it is still unknown under what initial conditions the formation of galaxies 
occurs.  A conventional idea is that the galaxies form at the local peaks of 
the initial density field. If this is really the case, then it is unlikely 
that the initial conditions establish any strong dependence on the 
environments since the gravitational collapse of the local density peaks 
must depend mostly on the mass enclosed rather than the environments around  
the peaks \citep{she-tor04}. 

However, there is another effect other than the local gravity that is 
responsible the formation of galaxies. At the initial stages, proto-galaxies 
interact with the surrounding matter, which originates the angular momentum.  
If the efficiency of the angular momentum generation by the external tidal 
field depends on the density of the environment, then the properties of 
galaxies like the morphology and the specific star formation rate may end up 
with having the environmental dependence. 

In this Letter, we construct a new theoretical model for the efficiency of 
the generation of the galaxy angular momentum as a function of the density 
of the environment, and show qualitatively how the environmental dependence 
of the galaxy properties should be at least partly established by the initial 
tidal condition.

\section{THE INITIAL TIDAL CONDITION}

In the tidal-torque scenario \citep{hoy49,pee69}, a proto-galaxy acquires 
the angular momentum ${\bf L} = (L_i)$ through tidal interaction with the 
surrounding matter.  The linear perturbation theory enables us to compute 
the angular momentum of a proto-galaxy in terms of Lagrangian quantities. 
According to this theory, if the shape of a proto-galaxy is not perfectly 
spherical, its angular momentum can be expressed as  
\citep{dor70,whi84,cat-the96} 
\begin{equation}
\label{eqn:ang}
L_{i} \propto \epsilon_{ijk}T_{jl}I_{lk}.
\end{equation}
This is the first-order approximation of the proto-galaxy angular momentum. 
Here, $\epsilon_{ijk}$ is an alternating symbol, and ${\bf I} = (I_{ij})$ 
and  ${\bf T} = (T_{ij})$ are the inertia momentum tensor of the proto-galaxy 
and the tidal shear tensor of the surrounding matter, respectively. The 
definitions of the two tensors are given, respectively as  
\begin{equation}
\label{eqn:it}
I_{ij} \equiv \int_{V_L} q_i q_j d^{3}q, \qquad 
T_{ij} \equiv \partial_{i}\partial_{j}\phi \bigg{|}_{{\bf q}_{cm}},
\end{equation}
where $V_L$ is the Lagrangian volume of a proto-galaxy,  ${\bf q} = (q_{i})$ 
is the Lagrangian position vector in the center-of-mass frame, ${\bf q}_{cm}$ 
is the Lagrangian center-of-mass of the proto-galaxy, and $\phi$ is the 
initial velocity potential, related to the linear density contrast 
$\delta \equiv (\rho-\bar{\rho})/\bar{\rho}$ as 
$\delta = \nabla^{2}_{\bf q}\phi({\bf q})$ 
($\bar{\rho}$ : the mean density of the background universe).

Since we consider the generation of the angular momentum on the proto-galactic 
mass scale, the density field $\delta$ is supposed to be averaged over the  
proto-galactic mass scale $M$. In other words, the linear density $\delta$ 
is not the actual one but the smoothed one by convolution with a filter on the 
proto-galaxy mass scale. Accordingly, the tidal tensor ${\bf T}$ in equation 
(\ref{eqn:it}) is a smoothed version of the actual tidal field on the mass 
scale $M$. In order to show explicitly that the tidal tensor in equation 
(\ref{eqn:it}) is smoothed on the proto-galaxy mass scale $M$, we use the 
superscripted notation, ${\bf T}^{M}$, from here on. The same notation rule 
applies to the inertia tensor ${\bf I}^{M}$ since it represents the
geometrical shape of the proto-galaxy, defined on the same mass scale $M$.  
Thus, equations (\ref{eqn:ang})-(\ref{eqn:it}) imply that the proto-galaxy 
angular momentum is generated by the tidal field smoothed on the same mass 
scale as the inertia tensor. 

Adopting a top-hat spherical filter whose functional form in the Fourier 
space is given as 
\begin{equation}
W(kR) = \frac{kR\sin(kR)-\cos(kR)}{kR},
\end{equation} 
one can relate the proto-galactic mass $M$ to the filtering radius $R$ as    
\begin{equation}
\label{eqn:mas}
M(R) = \frac{4\pi}{3}\bar{\rho}(1 + \delta)R^{3}. 
\end{equation} 
Here, the filtering radius $R$ represents the effective spherical radius 
(i.e., the linear size) of the proto-galaxy $R \propto V^{1/3}_{L}$. 

The tidal interaction of the proto-galaxy with the surrounding materials 
continues till the moment of the turn-around when $\delta \sim 1$.  
After the moment of the turn-around, the tidal influence from the 
surrounding matter effectively terminates, bringing the growth of the 
angular momentum to a halt. Thus, it is at the moment of the turn-around 
when the tidal effect of the surrounding matter is most influential. 
Although the nonlinear effect like galaxy-galaxy interaction might modify 
the galaxy angular momentum, it have been verified both numerically and 
observationally that the angular momentum of galaxies at present epoch still 
keeps quite well the initial memory of the surrounding matter distribution at 
the moment of the turn-around \citep{nav-etal04,tru-etal05}.

Now, let us express equation (\ref{eqn:ang}) in the principal axis frame 
of the tidal tensor ${\bf T}$: 
\begin{equation}
\label{eqn:pri}
L_{1} \propto (\lambda^{M}_{1} - \lambda^{M}_{2})I^{M}_{12},
\qquad
L_{2} \propto (\lambda^{M}_{2} - \lambda^{M}_{3})I^{M}_{23},
\qquad
L_{3} \propto (\lambda^{M}_{3} - \lambda^{M}_{1})I^{M}_{31},
\end{equation} 
where $\lambda^{M}_{1},\lambda^{M}_{2},\lambda^{M}_{3}$ are the three 
eigenvalues of the tidal tensor ${\bf T}^{M}$ in a decreasing order, and 
$I^{M}_{12},I^{M}_{23},I^{M}_{31}$ are the three off-diagonal components 
of the inertia tensor ${\bf I}^{M}$ in the principal axis frame of the tidal 
tensor ${\bf T}^{M}$.  

Equation (\ref{eqn:pri}) implies that only in case the principal axes of 
the inertia tensor ${\bf I}^{M}$ are not aligned with that of the tidal 
tensor ${\bf T}^{M}$, the angular momentum can be generated at first order. 
It also implies that the degree of the misalignment between the principal 
axes of ${\bf T}^{M}$ and ${\bf I}^{M}$ is directly proportional to the 
magnitude of the angular momentum. Recent N-body simulations demonstrated 
that the principal axes of the inertia tensor ${\bf I}^{M}$ are almost 
perfectly aligned with that of the tidal tensor ${\bf T}^{M}$, if the two 
tensors are smoothed at the same radius \citep{lee-pen00,por-etal02}. 

This empirical finding may be understood theoretically. In the Zel'dovich 
approximation \citep{zel70}, a proto-galaxy forms at the moment of 
shell-crossing when the principal axes of the tidal tensors are perfectly 
aligned with that of the inertia tensors. 

But, it is still possible for the galaxy angular momentum to be generated 
at first order. Although the tidal interaction is a {\it local} effect,  
the external tidal field has a {\it large-scale coherence} \citep{bon-etal96}. 
Thus, the tidal tensor smoothed at larger scales may be misaligned with the 
inertia tensor, and thus it can generate the galaxy angular momentum at first 
order. 

Let $\delta_{1}$ and $R_{1}$ be the linear density contrast and the 
effective spherical radius of the proto-galaxy of mass $M$. To consider 
the effect of the external tidal field smoothed at larger scales, we assume 
that the proto-galaxy is located in the environment whose linear density 
contrast and the scale radius are given as $\delta_{2} < \delta_{1}$ 
and $R_{2} > R_{1}$, respectively, satisfying a condition that the larger 
scale radius $R_{2}$ should enclose the same amount of mass, $M$.  
This condition guarantees that even though the external field is smoothed 
at larger scale $R_{2}$, it represents a local tidal effect acting on the 
proto-galactic mass $M$. Using the condition $M(R_{1}) = M(R_{2})$, we have 
\begin{equation}
\label{eqn:rad12}
R_{2} = R_{1}\left(\frac{1+\delta_{1}}{1+\delta_{2}}\right)^{1/3}.
\end{equation}
Since the values of $R_{1}$ and $\delta_{1}$ are fixed, equation 
(\ref{eqn:rad12}) implies that the lower the linear density $\delta_{2}$ is, 
the larger is the scale radius $R_{2}$. 

Now that the tidal tensor ${\bf T}^{M}$ is smoothed at larger scale $R_{2}$, 
it is no longer perfectly aligned with the inertia tensor ${\bf I}^{M}$ 
smoothed at scale $R_{1}$. The misalignment between ${\bf T}^{M(R_{2})}$ 
and ${\bf I}^{M(R_{1})}$ is expected to increase as the difference between 
$R_{1}$ and $R_{2}$ increases. To quantify the misalignment between 
${\bf T}^{M(R_{2})}$ and ${\bf I}^{M(R_{1})}$, we first rescale both the 
tidal and the inertia tensors to have unit magnitudes and become traceless as 
\begin{equation}
\label{eqn:hat}
\hat{\bf T}^{M(R_{2})} \equiv \frac{\tilde{\bf T}^{M(R_{2})}}
{\vert\tilde{\bf T}^{M(R_{2})}\vert}, \qquad 
\hat{\bf I}^{M(R_{1})} \equiv \frac{\tilde{\bf I}^{M(R_{1})}}
{\vert\tilde{\bf I}^{M(R_{1})}\vert},
\end{equation}
where $\vert\tilde{\bf T}^{M(R_{2})}\vert \equiv 
[\tilde{T}^{M(R_{2})}_{ij}\tilde{T}^{M(R_{2})}_{ij}]^{1/2}$ and 
$\vert\tilde{\bf I}^{M(R_{1})}\vert \equiv 
[\tilde{I}^{M(R_{1})}_{ij}\tilde{I}^{M(R_{1})}_{ij}]^{1/2}$, with 
$\tilde{\bf T} \equiv {\bf T} - {\rm Tr}({\bf T})/3$ and 
$\tilde{\bf I} \equiv {\bf I} - {\rm Tr}({\bf I})/3$.

By equation (\ref{eqn:ang}), the square of the magnitude of the proto-galaxy 
specific (per mass) angular momentum is proportional to 
\begin{equation}
\label{eqn:l2}
\langle L^{2} \rangle  \propto \epsilon_{iab}\epsilon_{iuv}
\langle \hat{T}^{M(R_{2})}_{ac}\hat{T}^{M(R_{2})}_{cb}\hat{I}^{M(R_{1})}_{uw}
\hat{I}^{M(R_{1})}_{wv} \rangle.
\end{equation}
Since the principal axes of the inertia and the tidal tensors smoothed at 
the same filtering radius are almost perfectly correlated, we make an 
approximation $\hat{\bf I}^{M(R_{1})} \approx  \hat{\bf T}^{M(R_{1})}$ 
in equation (\ref{eqn:l2}), and we define $\beta$ as 
\begin{equation}
\label{eqn:bet}
\beta \equiv \left[\epsilon_{iab}\epsilon_{iuv}\langle\hat{T}^{M(R_{2})}_{ac}
\hat{T}^{M(R_{2})}_{cb}\hat{T}^{M(R_{1})}_{uw}\hat{T}^{M(R_{1})}_{wv}
\rangle\right]^{1/2}. 
\end{equation}
The value of $\beta$  basically represents the {\it efficiency} of the 
generation of the galaxy angular momentum at first order. It increases as 
the degree of the misalignment between the principal axes of the tidal tensor 
${\bf T}^{M(R_{2})}$ and ${\bf T}^{M(R_{1})}$ increases.  Thus, it is 
expected that it increases as the density of the environment decreases 
according to equation (\ref{eqn:rad12}). 

To see this explicitly, we evaluate the value of $\beta$, using the following 
approximation whose validity was justified by the Monte-Carlo simulation 
\citep{lee-pen01}:
\begin{equation}
\label{eqn:app}
\langle\hat{T}^{M(R_{2})}_{ac}\hat{T}^{M(R_{2})}_{cb}\hat{T}^{M(R_{1})}_{uw}
\hat{T}^{M(R_{1})}_{wv}\rangle \approx
\frac{\langle\tilde{T}^{M(R_{2})}_{ac}\tilde{T}^{M(R_{2})}_{cb}
\tilde{T}^{M(R_{1})}_{uw}\tilde{T}^{M(R_{1})}_{wv}\rangle}
{\vert\tilde{\bf T}^{M(R_{2})}\vert^{2}\vert\tilde{\bf T}^{M(R_{1})}\vert^{2}}.
\end{equation}
By the Wick theorem, 
$\langle\tilde{T}^{M(R_{2})}_{ac}\tilde{T}^{M(R_{2})}_{cb}
\tilde{T}^{M(R_{1})}_{uw}\tilde{T}^{M(R_{1})}_{wv}\rangle$, equation 
(\ref{eqn:app}) is written as
\begin{eqnarray}
\label{eqn:wic}
\langle\tilde{T}^{M(R_{2})}_{ac}\tilde{T}^{M(R_{2})}_{cb}
\tilde{T}^{M(R_{1})}_{uw}\tilde{T}^{M(R_{1})}_{wv}\rangle &=& 
\langle\tilde{T}^{M(R_{2})}_{ac}\tilde{T}^{M(R_{2})}_{cb}\rangle
\langle\tilde{T}^{M(R_{1})}_{uw}\tilde{T}^{M(R_{1})}_{wv}\rangle \nonumber \\
&+&\langle\tilde{T}^{M(R_{2})}_{ac}\tilde{T}^{M(R_{1})}_{wv}\rangle
\langle\tilde{T}^{M(R_{1})}_{uw}\tilde{T}^{M(R_{2})}_{cb}\rangle \nonumber \\
&+&\langle\tilde{T}^{M(R_{2})}_{ac}\tilde{T}^{M(R_{1})}_{uw}\rangle
\langle\tilde{T}^{M(R_{1})}_{wv}\tilde{T}^{M(R_{2})}_{cb}\rangle.
\end{eqnarray}
The correlations between the components of the tidal tensors can be 
written as 
\begin{equation}
\label{eqn:cor}
\langle T^{M(R_1)}_{ij}T^{M(R_2)}_{kl}\rangle = 
\partial_{i}\partial_{j}\partial_{k}\partial_{l}{\bf \nabla}^{-4}
\int P(k)W_{TH}(k;R_{1})W_{TH}(k;R_{2})d^{3}k.
\end{equation}
Through equations (\ref{eqn:bet})-(\ref{eqn:cor}), it is straightforward 
to derive
\begin{equation}
\label{eqn:fin}
\beta = \left(1 - \frac{\sigma^{4}_{c}}{\sigma^{2}_{1}\sigma^{2}_{2}}
\right)^{1/2},
\end{equation}
where the three mass variances $\sigma^{2}_{1},\sigma^{2}_{2},\sigma^{2}_{c}$ 
are defined as 
\begin{eqnarray}
\label{eqn:sig1}
\sigma^{2}_{1} &=& \frac{1}{2\pi^2}\int P(k) W^{2}(kR_{1}) k^{2}dk, 
\\
\label{eqn:sig2}
\sigma^{2}_{2} &=& \frac{1}{2\pi^2}\int P(k) W^{2}(kR_{2}) k^{2}dk, 
\\
\label{eqn:sigc}
\sigma^{2}_{c} &=& \frac{1}{2\pi^2}\int P(k) W(kR_{1})W(kR_{2})k^{2}dk, 
\end{eqnarray}
with the linear power spectrum $P(k)$. 

\begin{figure}
\begin{center}
\plotone{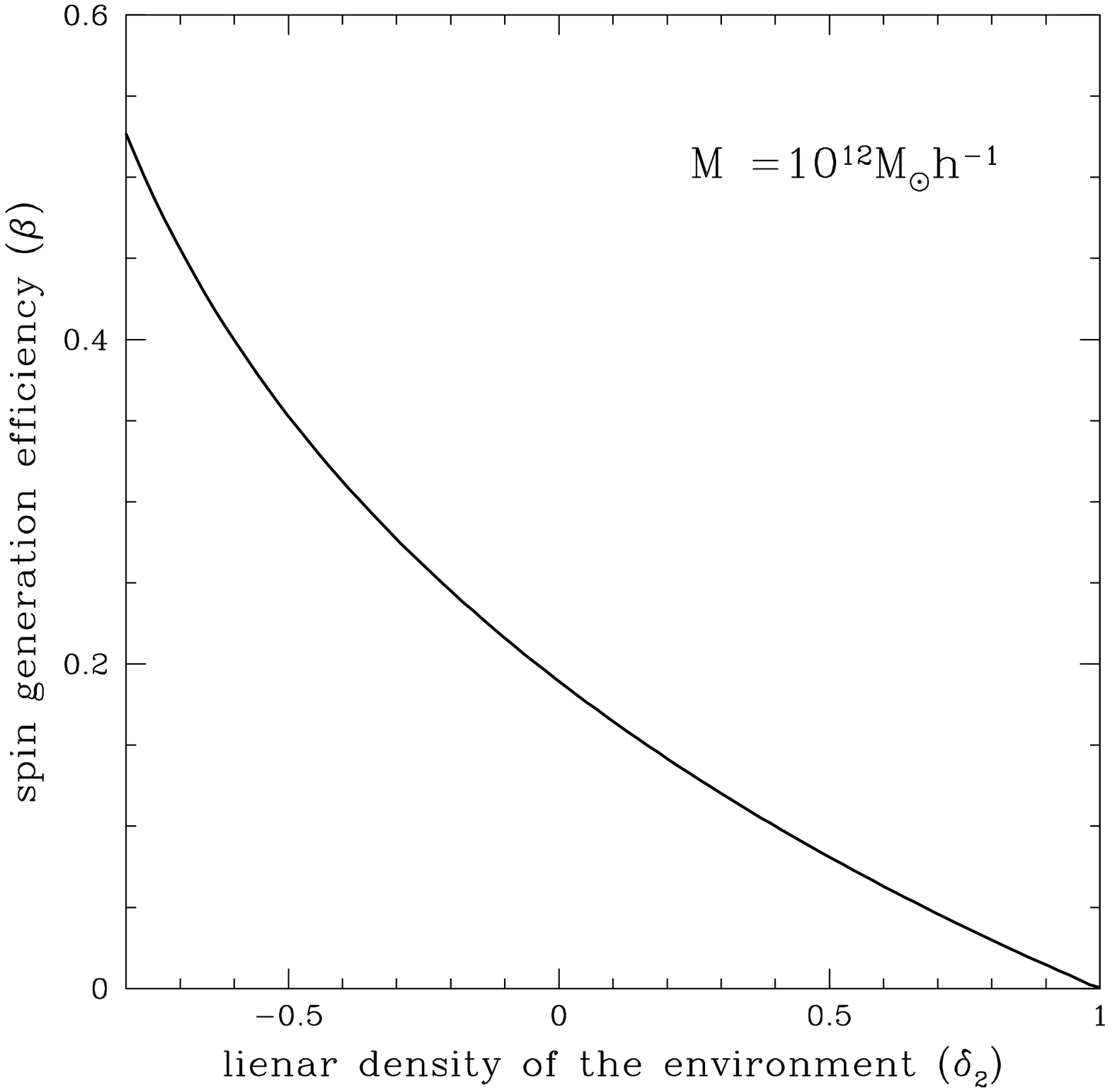}
\caption{Spin-generation efficiency of a galactic dark halo with mass 
$10^{12}h^{-1}M_{\odot}$ as a function of the linear density contrast 
of the environment where the galaxy is located.
\label{fig:ang}}
\end{center}
\end{figure}

Figure \ref{fig:ang} plots $\beta$ as a function of the linear density 
contrast of the environment $\delta_{2}$ through equations 
(\ref{eqn:rad12})-(\ref{eqn:sigc}). For $P(k)$ in equations 
(\ref{eqn:sig1})-(\ref{eqn:sig2}), we use the approximation given 
by \citet{bar-etal86} for the concordance cosmology with 
$\Omega_{m}=0.3,\Omega_{\Lambda}=0.7,h=0.7,\Omega_{b}=0.044$, and 
$\sigma_{8}=0.9$. For this plot, we use the typical galaxy mass scale 
$M = 10^{12}h^{-1}M_{\odot}$.  As seen clearly, $\beta$ decreases 
monotonically as $\delta_{2}$ increases, reducing to zero as the value of 
$\delta_{2}$ approaches $\delta_{1}$. This result shows how the generation 
of the galaxy angular momentum depend on the environment: high-angular 
momentum in the low-density region ($\delta_{2} << \delta_{1}$) and 
low-angular momentum in the high-density region 
($\delta_{2} \sim \delta_{1}$). It provides a theoretical explanation 
to the recent numerical finding that the spin parameters of void galaxies 
are higher than those of cluster galaxies \citep{avi-etal05}. 

\section{DISCUSSION AND CONCLUSION}

In the cold dark matter (CDM) paradigm, the basic mechanism for the star 
formation in galaxies is described as follows.  The hot gas falls into the 
gravitational potential well of the galactic dark halo, and goes through 
radiative cooling. A cloud of the cool gas in the halo potential well 
fragments into many clumps, and those clumps can further collapse via 
self-gravity to form stars.  The angular momentum of a galactic dark halo 
hinders the formation of stars in it: The high angular momentum causes high 
rotational speed of the gas clumps, which slows down the collapse of clumps 
into stars. 

In $\S 2$, it is shown that the galaxies in the low-density environment will 
end up with having high angular momentum. Consequently, it will take long time 
before stars form in the galaxies located in the low-density environment. 
For example, if a galaxy is located in the lowest-density environment like a 
void, then stars in the void galaxy must form only very recently. In other 
words, the void galaxies are predicted to have young stellar populations and 
show higher specific star formation rate at present epoch. 
One should note here that it is true only for those void galaxies whose 
sizes are large enough to hold gas.  If a void galaxy is small, 
then the UV-background will photo-evaporate all gas in the galaxy,  
suppressing any star-formation activity \citep[e.g.,][]{sus-ume04,hoe-etal05}. 

This theoretical insight agrees with the recent observational report 
that the large void galaxies at fixed luminosity and surface brightness 
profile are disk-like, having bluer colors and higher specific star 
formation rate than the wall galaxies,  which was not explained by 
the previous morphology-density relation \citep{vog-etal04}.

The properties of galaxies located in the high-density environment like 
galaxy clusters may be explained in the same line. A low angular momentum 
is expected for a proto-galaxy in a cluster due to the strong alignment 
between the inertia and the tidal tensors in the high-density. 
Thus, it will be quite faster to form stars in the cluster galaxies, 
because there is no hindrance caused by the galaxy angular momentum. 
However, for the cluster galaxies, it is expected that the evolutionary 
process like merging and interaction must play more significant role in 
establishing the environmental dependence. 

As a final conclusion, we have provided a new insight into the puzzle of 
galaxy formation by showing that the environmental dependence of galaxy 
properties has an aspect of destiny, being caused by the initial cosmological 
conditions.

\acknowledgments

J.L. acknowledges stimulating discussion with W. T. Kim. 
This work is supported by the research grant No. R01-2005-000-10610-0 from 
the Basic Research Program of the Korea Science and Engineering Foundation.


\clearpage

\end{document}